\documentclass[preprint,review,12pt, numbers,sort&compress]{elsarticle}

\usepackage{amssymb}
\usepackage{tabularx}
\usepackage{color}
\begin{document}

\begin{frontmatter}



\title{Broadband and Omnidirectional Anti-reflection Layer for III/V Multi-junction Solar Cells}


%%\author{}

%%\address{}

\author[FOM]{Silke L. Diedenhofen\corref{cor1}\fnref1{}}
\cortext[cor1]{Corresponding author}
\address[FOM]
{FOM Institute AMOLF, c/o Philips Research Laboratories, High-Tech
Campus 4, 5656 AE Eindhoven, The Netherlands}
\ead{diedenhofen@amolf.nl} \fntext[]{Present address: ICFO - The
Institute of Photonic Sciences, Av. Carl Friedrich Gauss, 3, 08860
Castelldefels (Barcelona), Spain.}

\author[FOM]{Grzegorz Grzela}
%\address[FOM Institute AMOLF] {FOM Institute AMOLF, c/o Philips
%Research, High-Tech Campus 4, 5656 AE Eindhoven}

\author[Nij]{Erik Haverkamp}
\address[Nij] {Radboud University Nijmegen, Institute for Molecules and
Materials, Applied Materials Science, Heyendaalseweg 135, 6525 AJ
Nijmegen, The Netherlands}

\author[Nij]{Gerard Bauhuis}
%\address[Radboud University Nijmegen] {Nijmegen}

\address[Phil] {Philips Research Laboratories,
High-Tech Campus 4, 5656 AE Eindhoven, The Netherlands}
\address[COBRA]{COBRA Research Institute, Eindhoven University of
Technology, P.O. Box 513, 5600 MB Eindhoven, The Netherlands}

\author[Nij]{John Schermer}
%\address[Radboud University Nijmegen] {Radboud University Nijmegen}

\author[FOM,COBRA]{Jaime G\'omez Rivas\corref{cor2}}
\cortext[cor2]{Principal corresponding author}
%\address[FOM Institute AMOLF] {FOM Institute AMOLF, c/o
%Philips Research, High-Tech Campus 4, 5656 AE Eindhoven}
%\address[Eindhoven University of Technology]{Applied Physics,
%Photonics \& Semiconductor Nanophysics, Eindhoven University of
%Technology, 5600 MB Eindhoven, The Netherlands}
\ead{rivas@amolf.nl}

\begin{abstract}
We report a novel graded refractive index antireflection coating
for III/V quadruple solar cells based on bottom-up grown tapered
GaP nanowires. We have calculated the photocurrent density of an
InGaP-GaAs-InGaAsP-InGaAs solar cell with a MgF$_2$/ZnS double
layer antireflection coating and with a graded refractive index
coating. The photocurrent density can be increased by 5.9~\% when
the solar cell is coated with a graded refractive index layer with
a thickness of 1 $\mu$m. We propose to realize such a graded
refractive index layer by growing tapered GaP nanowires on III/V
solar cells. For a first demonstration of the feasibility of the
growth of tapered nanowires on III/V solar cells, we have grown
tapered GaP nanowires on AlInP/GaAs substrates. We show
experimentally that the reflection from the nanowire coated
substrate is reduced and that the transmission into the substrate
is increased for a broad spectral and angular range.

\end{abstract}

\begin{keyword}

III/V Solar cells \sep anti-reflection \sep semiconductor nanowires


\end{keyword}

\end{frontmatter}

% \linenumbers

%%\section{}
%%\label{}

\section{Introduction}
Over the last years the photovoltaic research has brought many
solutions to increase the efficiency of solar cells. While triple
junction solar cells with efficiencies exceeding 40~\% have been
demonstrated \cite{King07,Guter09}, research is focusing now on
the realization of quadruple solar cells \cite{Stan10}. These
solar cells have expected efficiencies exceeding 50~\%
\cite{King07a}. However, for this type of solar cells, the
standard double layer antireflection coatings are not sufficient
for reducing the reflection over the broad wavelength range that
is absorbed by the multiple junctions \cite{Stan10,Friedman10}.
The standard double layer antireflection coatings typically have a
bandwidth lower than one octave, e.g., 400~nm to 700~nm or 800~nm
to 1100~nm \cite{Hobbs07}, while quadruple solar cells require an
antireflection coating operating from $\sim$350~nm to
$\sim$1700~nm \cite{King07a}. To approach an optimal performance
and the maximum efficiency that has been theoretically calculated
broad-band antireflection layers are necessary. The spectral range
of an antireflection coating is not the only factor limiting the
efficiency of multi-junction solar cells. As these solar cells
will be mainly installed in solar concentrator systems, the
angular response has to be optimized as well.

One way to achieve the desired broadband and omni-directional
antireflection layers is by employing the so-called moth-eye
concept \cite{Bernhard67}. In this concept, nanostructures forming
an effective medium result in a gradual increase of the refractive
index from that of air (n~=~1) to that of the solar cell medium
(n~$\sim$~3.3). Recently, different graded refractive index layers
have been demonstrated
\cite{Xi07,Huang07,Lee08,Diedenhofen09,Tommila10,
Leem2011,Kumar2011,Tseng2011,Baek2012}. These subwavelength
structures can be achieved in different ways, e.g., by etching
nanostructures into the surface
\cite{Huang07,Tommila10,Leem2011,Kumar2011,Tseng2011, Baek2012},
or by oblique-angle deposition \cite{Xi07}, or by chemical
deposition of nanostructures \cite{Lee08,Diedenhofen09}. When the
nanostructures are fabricated by etching, the material  of the
nanostructures is typically the same forming the active layer of
the solar cell. Deposition of nanostructures offers a wider range
of materials for the nanostructures, thus higher flexibility in
the design of antireflection coatings\cite{Xi07,Lee08,
Diedenhofen09}.

It has been recently demonstrated that graded refractive index
layers can be fabricated by bottom-up grown tapered GaP nanowires
or layers of nanowires with a broad distribution of nanowire
lengths \cite{Diedenhofen09}. The nanowires were grown on a GaP
substrate, and an increased transmission through the GaP substrate
with respect to a bare GaP substrate was demonstrated. The
bottom-up growth method used for fabricating nanowires allows
hetero-epitaxial growth of a passive antireflection layer on top
of the active layer of the solar cell. Having a passive
antireflection layer is advantageous because of the following
reasons: Surface carrier recombination can be dominant in active
nanostructured materials due to their large surface. This effect
is not relevant in the case of passive nanostructures, such as the
proposed antireflection nanowire layers, since light is not
absorbed in these structures. Carriers are only generated in the
active multi-junction solar cell, beneath the nanowire coating,
where they can be separated and extracted.

In this manuscript, we calculate the photocurrent densities of
III/V quadruple solar cells covered with different antireflection
coatings. These calculations reveal that a graded refractive index
layer increases the photocurrent density of the solar cell with
respect to a solar cell coated with a standard MgF$_2$/ZnS double
layer antireflection coating. We propose to realize such graded
refractive index layers with tapered GaP nanowires and demonstrate
the growth of tapered GaP nanowires on a layer of AlInP that is
lattice matched to GaAs. We use GaP nanowires because of the large
refractive index of GaP (n $\simeq$ 3.3) and its high electronic
bandgap energy ($\lambda_{\rm gap}$ = 549 nm). AlInP is used as
substrate for the nanowire growth because it is considered the
most suitable material to form the upper window layer in a
quadruple III/V solar cell structure \cite{Tommila10}.
Furthermore, we have measured the reflectance and transmittance of
an AlInP/GaAs substrate coated with nanowires and without
nanowires. We find that the layer of tapered nanowires reduces the
total reflectance from the AlInP/GaAs substrate for a broad range
of wavelengths and concomitantly increases the omni-directional
transmission into and absorption in the AlInP/GaAs substrate.
These measurements constitute a proof-of-principle as they are the
first experimental verification that tapered GaP nanowires could
be used as an antireflection layer for III/V solar cells.

\section{Anti-reflection Layers}
\begin{figure}[htp] \center
\includegraphics[width=140mm]{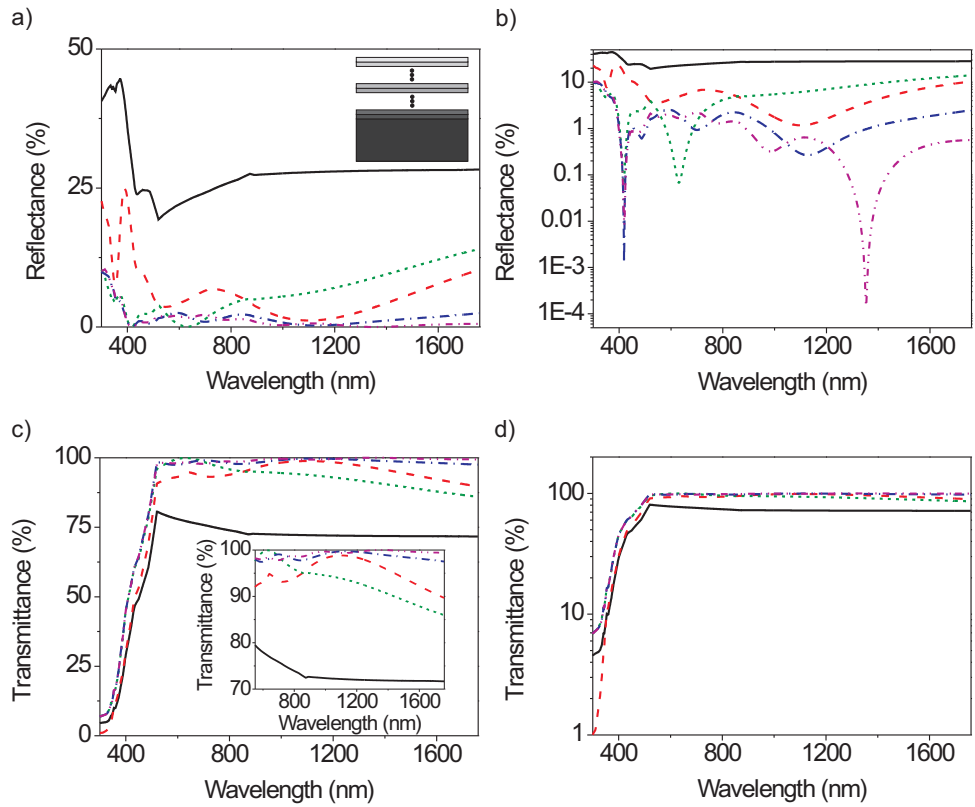}
\caption{a) and b) Calculation of the reflectance from and c) and
d) transmittance through a 30~nm thin AlInP layer coated with
different anti-reflection layers in linear scale (a,c) and
logarithmic scale (b,d). The black solid curves correspond to the
reflectance and transmittance of a bare AlInP layer on top of GaAs
substrate. The AlInP layer is coated with a MgF$_2$/ZnS
anti-reflection layer with thicknesses of 89~nm (ZnS) and 105~nm
(MgF$_2$) (red dashed curves), a parabolically increasing
refractive index layer from 1.0 to 3.3 with a thickness of 250~nm
(olive short-dashed curves), 500~nm (blue dash-dotted curves), and
1~$\mu$m (magenta dash-dot-dotted curves). The inset in a)
displays a schematic of the layout of the calculations in which a
change in color represents a change in refractive index. The inset
in c) shows the same calculations as in c) but in a narrower
wavelength range.} \label{FIGCalc}
\end{figure}

To show that the efficiency of a III/V quadruple solar cell can be
increased by adding a graded refractive index layer, we have
calculated the reflection from and transmission through an AlInP
window layer coated with different anti-reflection layers. For
these calculations the III/V solar cell was considered to consist
of InGaP, GaAs, InGaAsP, and InGaAs with electronic bandgaps that
correspond to cut-off wavelengths of respectively 649~nm, 873~nm,
1215~nm, and 1770~nm. The calculations are performed using the
transfer-matrix method \cite{Yeh48}. The refractive index of AlInP
varies from 3.06~-~2.14i at 350~nm to 2.77 at 1700~nm, and the
refractive index of GaAs from 3.28~-~2.06i at 350~nm to
$3.36-5.32\cdot10^{-5}$i at 1700~nm. These values have been
determined from ellipsometry measurements of the materials used in
the experiments shown in the next section. The refractive indices
of ZnS and MgF$_2$ vary between 2.8~-~0.13i and 2.27, and 1.38 and
1.37, respectively for wavelengths between 350~nm and 1700~nm.
Figure~\ref{FIGCalc} shows the calculated reflectance from an
AlInP layer with a thickness of 30~nm coated with different
anti-reflection layers in (a) linear and (b) logarithmic scale.
The black solid curves in Figures~\ref{FIGCalc}a and b show the
calculated reflectance of the bare AlInP layer without
antireflection coating. The reflectance from an AlInP layer coated
with a MgF$_2$/ZnS double layer antireflection coating with
thicknesses of 89~nm (ZnS) and 105~nm (MgF$_2$) is displayed by
the red long-dashed curves in Figures~\ref{FIGCalc}a and b. The
calculated reflectance from that coating is lower than 10 \% for
wavelengths between 450~nm and 1730~nm. While this anti-reflection
coating reduces the reflection for a rather broad wavelength
range, the bandwidth is not optimal for reducing the reflectance
over the wavelength range in which III/V multi-junction solar
cells absorb light.

To show that the bandwidth of low reflectance can be increased by a
graded refractive index layer, we have calculated the reflectance
from a 30~nm thin AlInP layer that is coated with three different
lossless graded refractive index layers. A schematic of the layout
used in the calculations is given in the inset of
Figure~\ref{FIGCalc}a, in which a change in color represents a
change in refractive index. The refractive index of these layers is
parabolically increasing from 1.0 to 3.3, and the thicknesses of the
layers are 250~nm, 500~nm, and 1000~nm, respectively. In the
transfer-matrix calculation, the anti-reflection layer is
implemented by considering sublayers with a thickness of 10~nm and
the refractive index is increased parabolically from one layer to
the next. These calculations are displayed in Figures~\ref{FIGCalc}a
and b by the olive-short-dashed, the blue dash-dotted, and the
magenta dashed-dot-dotted curves, respectively. While the graded
refractive index layer with a thickness of 250~nm does not decrease
the reflectance significantly for wavelengths longer than 850~nm,
the reflectance of the graded refractive index layers with a
thickness of 500~nm and 1000~nm are around 1 \% or lower for a broad
wavelength range. The reflection is the lowest, when a 1000~nm thick
graded refractive index antireflection coating is applied to the
multi-junction solar cell. In this coating, the gradient of the
refractive index between adjacent sublayers is the smallest.

Corresponding to the reflectance calculations, we have performed
calculations of the transmittance through a 30~nm thin AlInP layer
coated with the same anti-reflection layers as above. The calculated
transmittance is shown in Figures~\ref{FIGCalc}c and d in linear and
logarithmic scale, respectively. From the calculated transmittance,
we have determined the solar spectrum weighted transmittance $SSWT$
of the different layers by

\begin{equation}\label{SolarWeightedTransmittance}
SSWT = \frac{\int F(\lambda)\cdot T(\lambda)d\lambda}{\int
F(\lambda)d\lambda},
\end{equation}

where $F(\lambda)$ and $T(\lambda)$ are the photon flux (AM1.5G,
1000 W/m$^2$) and the calculated transmittance, respectively. The
integration is performed in the wavelength range from 300~nm to
1760~nm. The solar weighted transmittances of the 5 different
samples are 69.2~\% for a bare solar cell, 86.3~\% for the
MgF$_2$/ZnS antireflection coating, 89~\% for the 250~nm thick
graded refractive index layer, 91.5~\% for the 500~nm thick graded
refractive index layer, and 91.7~\% for the 1~$\mu$m thick graded
refractive index layer. Therefore, the solar weighted
transmittance can be increased by 6.3~\% when the 1~$\mu$m thick
graded refractive index layer is used as an antireflection layer
instead of the double layer MgF$_2$/ZnS coating.

According to the calculations of reflectance from and
transmittance through the AlInP layer, we have determined the
reflectance from and transmittance through each individual layer
of a III/V multi-junction solar cell. We have modeled the optical
multilayer structure using the transfer-matrix method
\cite{Haverkamp05}. With this method, we have calculated the
reflectance and transmittance at each interface. From these
transmittances and reflectances, we have determined the absorption
in each sublayer. Multiplying the absorption in each subcell by
the photon flux results in the amount of photons absorbed in each
subcell. The internal quantum efficiency (IQE) of each sub cell is
set to unity, meaning that every photon absorbed in one of the
layers contributes to the photocurrent density of the cell. With
this IQE the electron-hole pair generation is calculated,
resulting in a sub-cell specific photocurrent. This method allows
to calculate an unlimited amount of optical layers but it is
limited by the difficulties in predicting the reflections between
sub-cells. The model has been verified by comparing the calculated
reflection of a single junction solar cell coated with different
antireflection layers to the measured reflection
\cite{Haverkamp05}. The photocurrent densities in each subcell and
the resulting photocurrent density of the quadruple solar cell are
given in Table~\ref{TABEfficiency} for a solar cell coated with a
perfect antireflection coating, without antireflection coating,
with a MgF$_2$/ZnS double-layer antireflection coating, and with
graded refractive index layers with parabolically increasing
refractive index from 1.0 to 3.3 over a thickness of 250~nm,
500~nm, and 1000~nm. The first column in Table~\ref{TABEfficiency}
shows the maximum photocurrent densities and the maximum
efficiency of the solar cell if a perfect anti-reflection layer is
used, i.e., all incident light within the absorption band of the
solar cell is transmitted into and absorbed in the solar cell. The
maximum photocurrent density that can be achieved when all
incident light is absorbed is 13.36 mA/cm$^2$. When applying a
graded refractive index layer with a thickness of 1000~nm and a
parabolically increasing refractive index from 1 to 3.3 to the
solar cell, a photocurrent density of 13.28 mA/cm$^2$ can be
achieved. This photocurrent density is limited by the bottom cell.
The calculated photocurrent density in the solar cell coated with
the MgF$_2$/ZnS double layer antireflection coating is 12.54
mA/cm$^2$ and is in all subcells lower than the photocurrent
density in the solar cell coated with the thickest graded
refractive index layer. The photocurrent density in the solar cell
coated with this graded refractive index layer is increased by
5.9~\% with respect to that of the cell with the double layer
antireflection coating.

\section{Tapered GaP Nanowires}

\begin{figure}[htp] \center
\includegraphics[width=90mm]{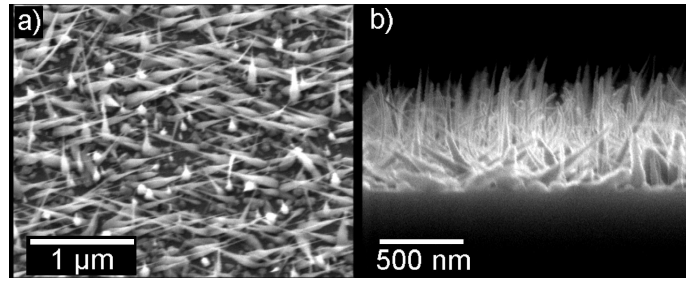}
\caption{a) Top-view and b) cross-sectional scanning electron
micrographs of GaP nanowires grown on AlInP.}\label{FIGSEM}
\end{figure}

To investigate the feasibility of nanowire graded refractive index
layers on top of multijunction solar cells, we have grown layers
of tapered GaP nanowires on top of AlInP using the
vapor-liquid-solid (VLS) growth mechanism \cite{Wagner64} by
metal-organic vapor phase epitaxy (MOVPE). AlInP is commonly used
as upper window layer in III/V solar cells \cite{Tommila10}. For
matching the growth conditions to that of a real solar cell
device, we have grown 30~nm of AlInP on top of a polished (100)
n$^{++}$-doped GaAs wafer. This AlInP layer has been protected by
a 300~nm thick GaAs layer. Before nanowire growth, we have removed
this GaAs layer by etching the substrate in 28\% NH$_4$OH :
H$_2$O$_2$ : H$_2$O (1 : 1 : 100). Immediately after this etching
step, gold with an equivalent layer thickness of 0.3~nm has been
evaporated and the gold coated substrate has been placed in the
MOVPE reactor. The GaP nanowires are grown for 180 seconds at a
temperature of 570~$^\circ$C using trimethyl-gallium (TMG) and
phosphine (PH$_3$) as precursors. A top-view scanning electron
micrograph of the GaP nanowires grown on AlInP is shown in
Figure~\ref{FIGSEM}a. The nanowires are grown epitaxially with
preferential growth in the $<$111$>$ direction, however, some
nanowires also grow normal to the substrate. From the
cross-sectional scanning electron micrograph in
Figure~\ref{FIGSEM}b, the thickness of the nanowire layer can be
determined to be 600~$\pm$~100~nm. The variation of the layer
thickness can be attributed to nanowires growing into different
crystallographic directions with different growth rates.

\begin{figure}[htp] \center
\includegraphics[width=140mm]{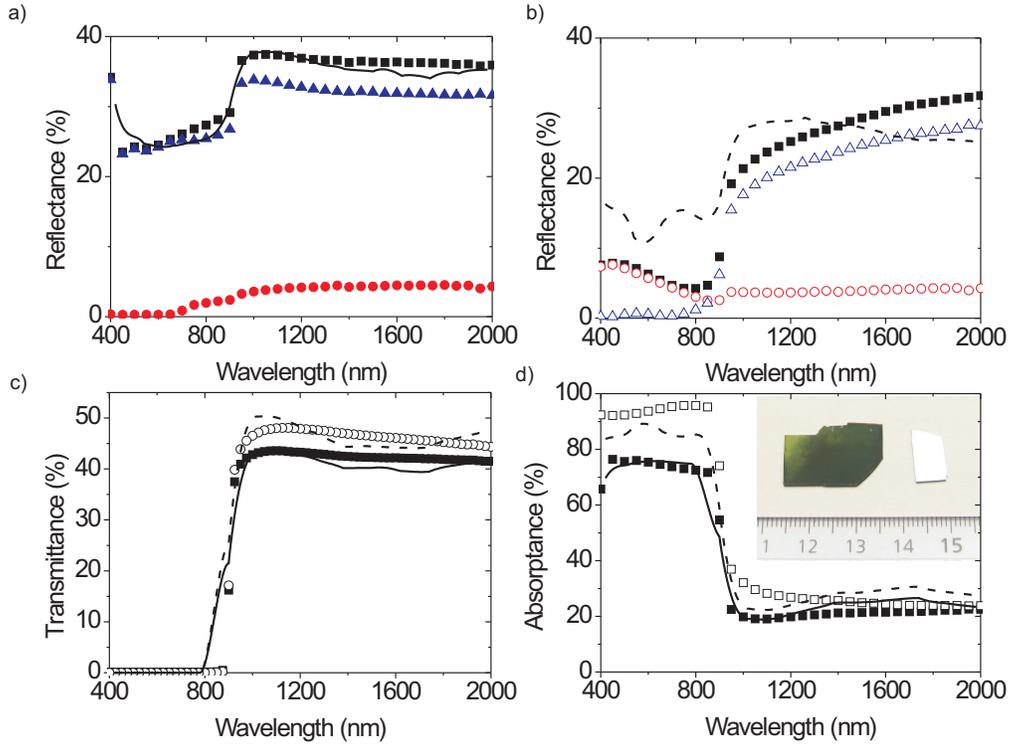}
\caption{a) Measured reflectance from the AlInP/GaAs substrate and
b) measured reflectance from the AlInP/GaAs substrate coated with
tapered nanowires. In both graphs the total reflectance (black
squares), specular reflectance (blue triangles), and diffuse
reflectance (red circles) are displayed. c) Measured total
transmittance through and d) measured absorptance in the AlInP/GaAs
substrate (solid squares) and the AlInP/GaAs substrate coated with
tapered nanowires (open squares). The solid curves correspond to the
calculated reflectance, transmittance, and absorptance of the
reference sample, and the dashed curves to the calculations of the
nanowire sample. The inset in d) shows a photograph of the
AlInP/GaAs substrate coated with nanowires and without
nanowires.}\label{FIGTotal}
\end{figure}

To determine the effect of the tapered nanowires on the reflection
and transmission, we have measured the total reflectance and the
diffuse reflectance from the AlInP/GaAs substrate and the
AlInP/GaAs substrate coated with tapered nanowires. We have
measured the total reflectance from both samples using a Lambda
950 spectrometer (PerkinElmer) consisting of a tungsten-halogen
and a deuterium lamp, an integrating sphere, and a photomultiplier
tube. For the reflectance measurements, the samples have been
mounted at the backside of the integrating sphere. The angle of
incidence was 8$^\circ$, allowing the measurement of the total and
diffusely reflected light. The diffusely reflected light is
measured by making a small opening in the integrating sphere in
such a way that the specularly reflected light escapes from the
sphere. The reflection measurements on the samples are normalized
by the reflection of a white-standard to obtain the absolute
reflectance of the array. From the total reflectance, $R$, and the
diffuse reflectance, $R_{\rm dif}$, the specular reflectance,
$R_{\rm spec}$, can be determined by $R_{\rm spec}=R-R_{\rm dif}$.
Figure~\ref{FIGTotal}a displays the measured total reflectance
(black squares), diffuse reflectance (red circles), and the
resulting specular reflectance (blue triangles) from the
AlInP/GaAs substrate. The total and specular reflectance is
similar (28~\%) at wavelengths below the electronic bandgap of
GaAs (865~nm), and the diffuse reflectance is negligible. This low
diffuse reflectance is expected for a flat layer. For wavelengths
longer than 865~nm, the reflectance increases to 36~\%, as the
reflectance of the backside of the substrate is contributing to
the measurement. At these wavelengths, the diffuse reflectance is
slightly increased, indicating that the backside of the sample is
rougher than the front side. From these two measurements, we can
understand the contribution to the reflection of the substrate
backside and the AlInP front side of the sample. The black solid
curve in Figure 3a shows a calculation of the specular reflection
of an AlInP/GaAs layer for comparison done with the transfer
matrix method. For this calculation, we have slightly varied the
refractive indices of AlInP and GaAs for fitting the measurements.
The resulting refractive index of AlInP varies from $n = 3.82 -
0.24i$ at 400~nm to 3.04 at 2000 nm, the refractive index of GaAs
varies from $n = 3.86 - 2.27i$ at 400~nm to $3.15 - 10^{-4}i$ at
2000~nm. These refractive indices are the only fitting parameters
in the determination of the reflection.

In a next step, we have measured the total and diffuse reflectance
from the AlInP/GaAs substrate coated with tapered nanowires. These
measurements are displayed in Figure~\ref{FIGTotal}b. The total
reflectance (black squares) is strongly reduced compared to the
measurement of the AlInP/GaAs substrate. While the total
reflectance is lower than 10~\% at wavelengths below 873~nm, the
total reflectance increases for wavelengths above the electronic
bandgap of GaAs as in the case of the bare AlInP/GaAs substrate.
This increased total reflectance can be attributed to the
reflectance of the backside of the substrate. The diffuse
reflectance (red circles) is dominating the total reflectance for
wavelengths shorter than 873~nm. The increasing diffuse
reflectance with decreasing wavelength is expected and due to
scattering of light from the thick bottom part of the nanowires
\cite{Muskens08}. For wavelengths longer than 865~nm, the diffuse
reflectance is similar to that of the AlInP/GaAs substrate,
indicating that scattering of light from the nanowires is
negligible at these wavelengths.

Although these first measurements demonstrate that the reflectance
of the AlInP/GaAs substrate coated with tapered nanowires is
significantly reduced with respect to the bare AlInP/GaAs
substrate, the reflectance of the nanowire sample needs to be
further decreased for a perfect anti-reflection coating.
Therefore, a smaller gradient of the refractive index has to be
realized. Further, the bottom diameter of the nanowires has to be
decreased to reduce the diffuse reflection at short wavelengths
and the density of nanowires has to be increased to achieve a high
index of refraction at the interface with the AlInP. The dashed
curve in Figure 3b shows a fit to the measurements using the
transfer matrix method. The spatial dependence of the refractive
index was used as a fitting parameter, obtaining a parabolic
increase from 1.1 at the air-nanowire interface to 1.4 at the
nanowire-substrate interface. While the fit describes the
measurement quantitatively at wavelengths longer than 870 nm,  it
is limited at visible wavelengths due to the scattering of light
by the nanowires. We consider the layer homogeneous for the fit,
neglecting this scattering in the calculation. Due to the
scattering, the Fabry-P\'erot oscillations that are visible in the
calculation are not present in the measurements. Further, the fit
overestimates the reflectance as light is preferentially scattered
into the substrate with a high refractive index
\cite{Catchpole08}.

To determine the enhancement of the coupling of light into the
AlInP/GaAs substrate by the tapered nanowires, we have measured
the total transmittance through the AlInP/GaAs substrate (solid
squares in Figure~\ref{FIGTotal}c) and through the AlInP/GaAs
substrate coated with tapered nanowires (open squares). The solid
and dashed curves in Figure 3c show the calculated transmittance
through the AlInP/GaAs substrate and the nanowire sample,
respectively. For wavelengths below 873~nm, the transmission is
negligible in both measurements, as the light gets absorbed in the
360~$\mu$m thick GaAs substrate. For wavelengths above the
electronic bandgap of GaAs (865~nm), the transmittance through the
bare AlInP/GaAs substrate increases strongly, reaching 43~\% for
wavelengths longer than 950~nm. The transmittance through the
AlInP/GaAs substrate coated with tapered nanowires is 47~\%. This
transmittance in both measurements is influenced by residual
free-carrier absorption for wavelengths above the electronic
bandgap of GaAs, which also causes the rather low increase of the
transmittance through the AlInP/GaAs substrate coated with
nanowires with respect to the bare AlInP/GaAs substrate. This
residual absorptance $A$ can be determined by $A = 100\%-R-T$,
with $R$ and $T$ being the total reflectance and transmittance,
respectively. The absorptance in the AlInP/GaAs substrate and in
the AlInP/GaAs substrate coated with tapered nanowires is
displayed in Figure~\ref{FIGTotal}d. The absorption in the
nanowire coated sample (open squares) is increased with respect to
the uncoated substrate (solid squares) for all wavelengths. For
the wavelength range below 873~nm, the fact that less light is
reflected is the main cause for the enhanced absorption
(absorptance higher than 90\%) in the sample coated with
nanowires. For the wavelength range above 873~nm, the increased
absorption might be attributed to diffusely transmitted light due
to the nanowires. The diffusely transmitted light travels through
the GaAs with a longer optical path length, resulting in a higher
absorption probability due to free-carrier absorption. The solid
and dashed curves in Figure~\ref{FIGTotal}d show the calculated
absorptance of the AlInP/GaAs substrate and the nanowire sample,
respectively, using the transfer matrix method. According to the
calculation of the reflectance of the nanowire sample, the
calculated absorptance in the nanowire sample is underestimated.
The inset in Figure~\ref{FIGTotal}d shows a photograph of the
AlInP/GaAs substrate coated with nanowires (sample on the left)
and without nanowires (sample on the right). The bare substrate is
shiny silverish, as expected from GaAs. The sample coated with
nanowires is black, demonstrating the high absorption in the
sample.

Analog to the calculations performed for determining the
photocurrent of the solar cells displayed in
Table~\ref{TABEfficiency}, we have calculated the photocurrent of
a solar cell coated with an antireflection layer that is gradually
increased from 1.1 to 1.4 over a thickness of 600~nm, resembling
our measurements. A solar cell coated with such an antireflection
layer would have a total current of 10.98 mA/cm$^2$. This total
current is increased by 1.5 mA/cm$^2$ compared to a bare cell.
However, the calculation only represents a lower limit of the
increase in photocurrent, as the absorptance in the solar cell
obtained from the calculation is underestimated with respect to
the expected absorption from the experiments. We note that while
the estimated photocurrent is lower than the photocurrent of a
solar cell coated with a double-layer antireflection coating, the
graded refractive index layer can be significantly improved by
growing a nanowire layer with a higher material filling fraction
at the bottom.

\begin{figure}[htp] \center
\includegraphics[width=90mm]{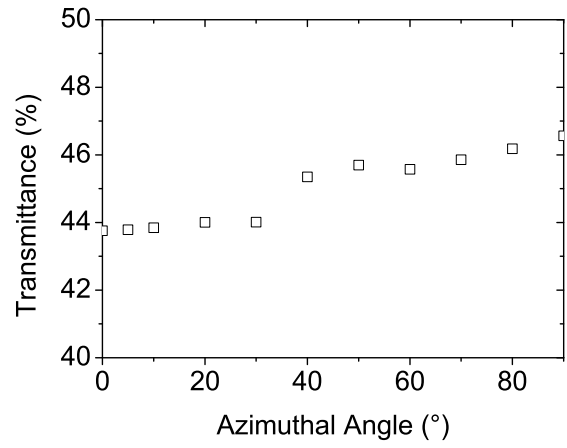}
\caption{Measured transmittance through the AlInP/GaAs substrate
coated with tapered nanowires as a function of azimuthal angle at a
wavelength of 1100~nm.}\label{FIGTransAzimuth}
\end{figure}

Because the tapered nanowires are grown preferentially into the
$<111>$B direction, we could expect that the transmittance through
the sample is anisotropic, i.e., the transmittance depends on the
azimuthal angle of the sample. To determine this anisotropy, we
have measured the zero-order transmission through the AlInP/GaAs
substrate coated with the tapered nanowires for different
azimuthal angles. Figure~\ref{FIGTransAzimuth} displays the
transmittance measured at 1100~nm as a function of azimuthal
angle. The rather constant transmittance with respect to this
angle indicates that the preferential growth along the $<111>$B
direction does not give rise to an anisotropic response. The
transmittance varies between 43.7~\% and 46.6~\%. These values are
slightly lower than the total transmission, indicating that a
small fraction of the intensity is diffusely scattered.
Importantly, the specular transmittance is for all azimuthal
angles higher than that measured through the bare AlInP/GaAs
substrate, which is 38.8~\% at 1100~nm.

\begin{figure}[htp] \center
\includegraphics[width=90mm]{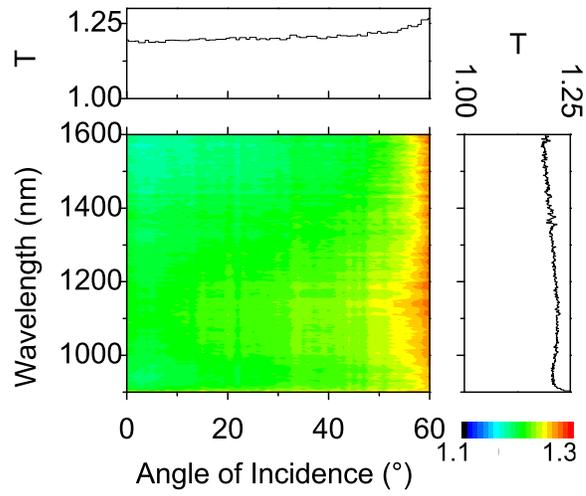}
\caption{Measured transmission through the AlInP/GaAs substrate
coated with nanowires normalized to the transmission through an
AlInP/GaAs substrate as a function of angle of incidence and
wavelength. The top inset shows a cut to the measurement at
1400~nm and the right inset shows a cut at an angle of incidence
of 30$^\circ$.}\label{FIGTransOmni}
\end{figure}

As multi-junction solar cells are usually installed in
concentrator systems, the desired anti-reflection coating should
omnidirectionally reduce the reflection and increase the
transmission into the solar cell. To determine the
omnidirectionality of the response of the antireflection layer, we
have measured the angle dependent zero-order transmission through
the AlInP/GaAs substrate coated with tapered nanowires and
normalized it to the transmission through the AlInP/GaAs substrate
for unpolarized light. This measurement is displayed in
Figure~\ref{FIGTransOmni}. For all wavelengths and angles up to
60$^\circ$, the normalized transmission is above unity, meaning
that for all wavelengths and angles, the transmission into the
AlInP/GaAs layer is increased by the presence of the nanowire
layer. The insets in Figure~\ref{FIGTransOmni} show cuts to the
measurement at a wavelength of 1400~nm as a function of angle of
incidence, and at an angle of incidence of 30$^\circ$ as a
function of wavelength. These cuts show that overall the
transmission is enlarged by around 20~\% and it further increased
to 25~\% for angles of incidence larger than 50$^\circ$. In a
Fresnel lens based concentrator system the light is typically
incident over angles from 0$^\circ$ to 30$^\circ$, however, higher
angles of incidence would allow thinner concentrator modules.

To determine if the nanowire layer is robust to the high
irradiance level in concentrator systems, we have illuminated the
sample with 125 suns (AM1.5G) for 4 hours. During the
illumination, the sample was mounted on a Cu base for temperature
control and the temperature of the sample was kept at
50~$^\circ$C. We have measured the reflection before and after
illumination on 5 different positions of the sample, and found
that there is no significant difference between the measurements.
From the reflection measurements before and after illumination
with 125 suns, we can conclude that the nanowire layer is stable
under concentrated sunlight.

\section{Conclusions}
In conclusion, we have presented calculations demonstrating that
the photocurrent density of a III/V quadruple solar cell can be
increased when applying a graded refractive index coating to the
solar cell instead of the standard MgF$_2$/ZnS double layer
antireflection coating. For demonstrating the feasibility of
graded refractive index layers on solar cells, we have grown
layers of tapered GaP nanowires on AlInP/GaAs substrates and we
have investigated the transmission and reflection properties of
these layers. We have shown that the reflection from AlInP/GaAs
substrates can be reduced by coating the substrate with a graded
refractive index layer. Our experiments reveal that the total
reflection is indeed reduced over a broad spectral range when
applying tapered GaP nanowires on top of an AlInP/GaAs substrate
and that the transmission into the substrate is increased for a
wide spectral and angular range. These results render tapered
nanowires a promising candidate for increasing the efficiency of
III/V multi-junction solar cells.

\section*{Acknowledgement}

We thank G. Immink for the growth of the nanowires, E. van Thiel
for technical assistance, J. Hoppenbrouwer for SEM analysis, and
Peter Thijs and Thierry de Smet for useful discussions. This work
is part of the research program of the "Stichting voor
Fundamenteel Onderzoek der Materie (FOM)", which is financially
supported by the "Nederlandse organisatie voor Wetenschappelijk
Onderzoek (NWO)" and is part of an industrial partnership program
between Philips and FOM.

\clearpage

\clearpage
\begin{table}[htbp]
    \centering
    \caption{Calculated photocurrent densities of a III/V quadruple junction solar cell coated with a perfect antireflection layer (AM 1.5G), a bare cell without antireflection layer, a MgF$_2$/ZnS double layer antireflection coating with thicknesses of 89~nm (ZnS) and 105~nm (MgF$_2$) (AR layer), and different graded refractive index (GI) anti-reflection layers that have a parabolically increasing refractive index from 1.0 to 3.3 over the given thickness.}
    \label{TABEfficiency}
\begin{tabularx}{15cm} {lXXXXXX}
\hline  & AM1.5G &
Bare cell & AR layer & 250 nm GI layer & 500 nm GI layer & 1000 nm GI layer  \\
\hline\hline
Top cell current [mA/cm$^2$]& 16.71  & 10.83 & 12.54 & 13.81 & 13.8 & 13.82 \\
Second cell current [mA/cm$^2$]& 15.4  & 11.45 & 14.4 & 14.88 & 15.09 & 15.08 \\
Third cell current [mA/cm$^2$] & 14.98  & 10.86 & 14.72 & 14.16 & 14.9 & 14.94  \\
Bottom cell current [mA/cm$^2$] & 13.36 & 9.57 & 12.54 & 11.91 & 13.13 & 13.28  \\
Resulting current [mA/cm$^2$] & 13.36   & 9.57 & 12.54 & 11.91 & 13.13 & 13.28  \\
\hline
\end{tabularx}
\end{table}
\clearpage

\clearpage

\clearpage

\clearpage

\clearpage

\clearpage



%% References


\bibliographystyle{elsarticle-num}
\bibliography{NWSolarCell}






% \end{thebibliography}

\end{document}